\documentclass{cmspaper}
\usepackage{graphicx,epsfig,rotating}
\usepackage{amssymb}
\usepackage[hyperindex]{hyperref}
\begin{document}

\newcommand{\ra}{\rightarrow}
\newcommand{\lsim}{\mbox{\raisebox{-.6ex}{~$\stackrel{<}{\sim}$~}}}
\newcommand{\gsim}{\mbox{\raisebox{-.6ex}{~$\stackrel{>}{\sim}$~}}}

  \begin{titlepage}

   \cmsnote{2002/024}
   \date{25 May 2002}

\title{ \boldmath{Prospects of Detecting Massive Charged Higgs from Hadronic Decay $H^{\pm} \rightarrow tb$ in CMS}}

  \begin{Authlist}

     P. ~Salmi and R.~Kinnunen 
     \Instfoot{HIP}{Helsinki Institute of Physics, Helsinki, Finland}

     N. ~Stepanov\Aref{a}
       \Instfoot{CERN}{European Laboratory for Particle Physics (CERN)}

  \end{Authlist}

\Anotfoot{a}{On leave from ITEP, Moscow, Russia}



  \begin{abstract}

The possibility to detect the massive charged Higgs boson $H^{\pm}$ using the hadronic decay channel $H^{\pm} \rightarrow tb$ in the associated production 
$pp \rightarrow tH^{\pm}+X$ in the CMS experiment at LHC is studied. There is a large background from $t\bar{t}b\bar{b}$ events which makes the observation difficult. Detection of a Higgs signal in this channel requires an excellent b-tagging performance. Good calorimeter mass resolution is also necessary for the full event reconstruction.
   
  \end{abstract}

  
  \end{titlepage}


  \section{Introduction}

The Minimal Supersymmetric Standard Model (MSSM) predicts five physical states
in the Higgs sector: three neutral scalars $h^{0}$, $H^{0}$, $A^{0}$ and a
charge conjugated pair $H^{\pm}$~\cite{1}. 
At the tree-level just two free parameters are enough
to fix masses of all these states~\cite{2}. The conventional choice is to use the mass
of CP-odd neutral Higgs, $m_{A}$, and the ratio of the vacuum expectations 
values of Higgs doublets, $\tan\beta$. For the values of $\tan\beta$ considered
here, the mass of $H^{\pm}$ does not differ considerably from that of $A^{0}$.

\begin{figure}[hbtp]
\begin{center}
\epsfig{file=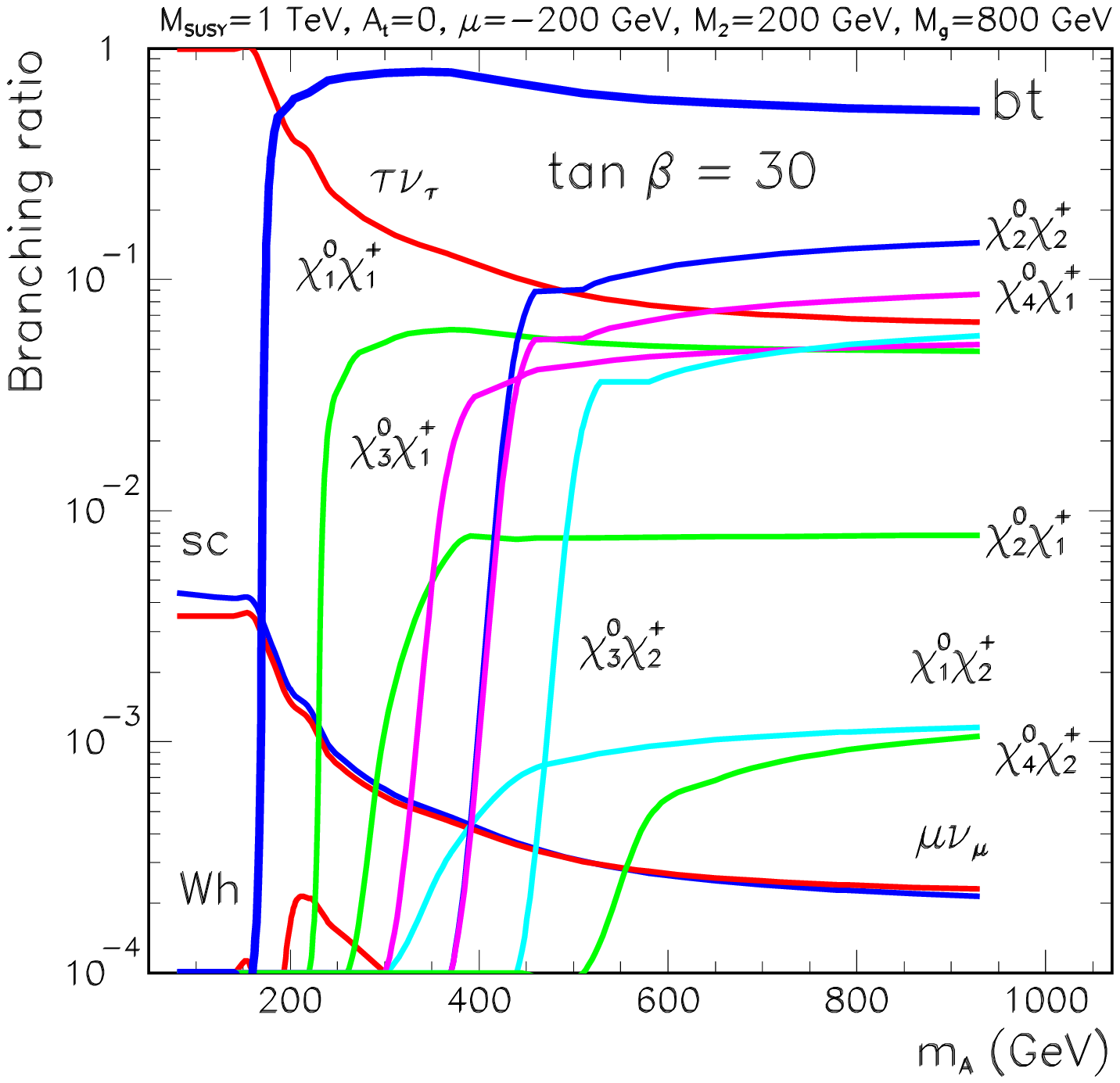,width=.65\linewidth}
\caption{The branching ratios of $H^{\pm}$ decay for $\tan \beta = 30$.}
\vspace{9mm}
\label{fig:beta30br}  
\end{center}
\end{figure}

\begin{figure}[hbtp]
\begin{center}
\epsfig{file=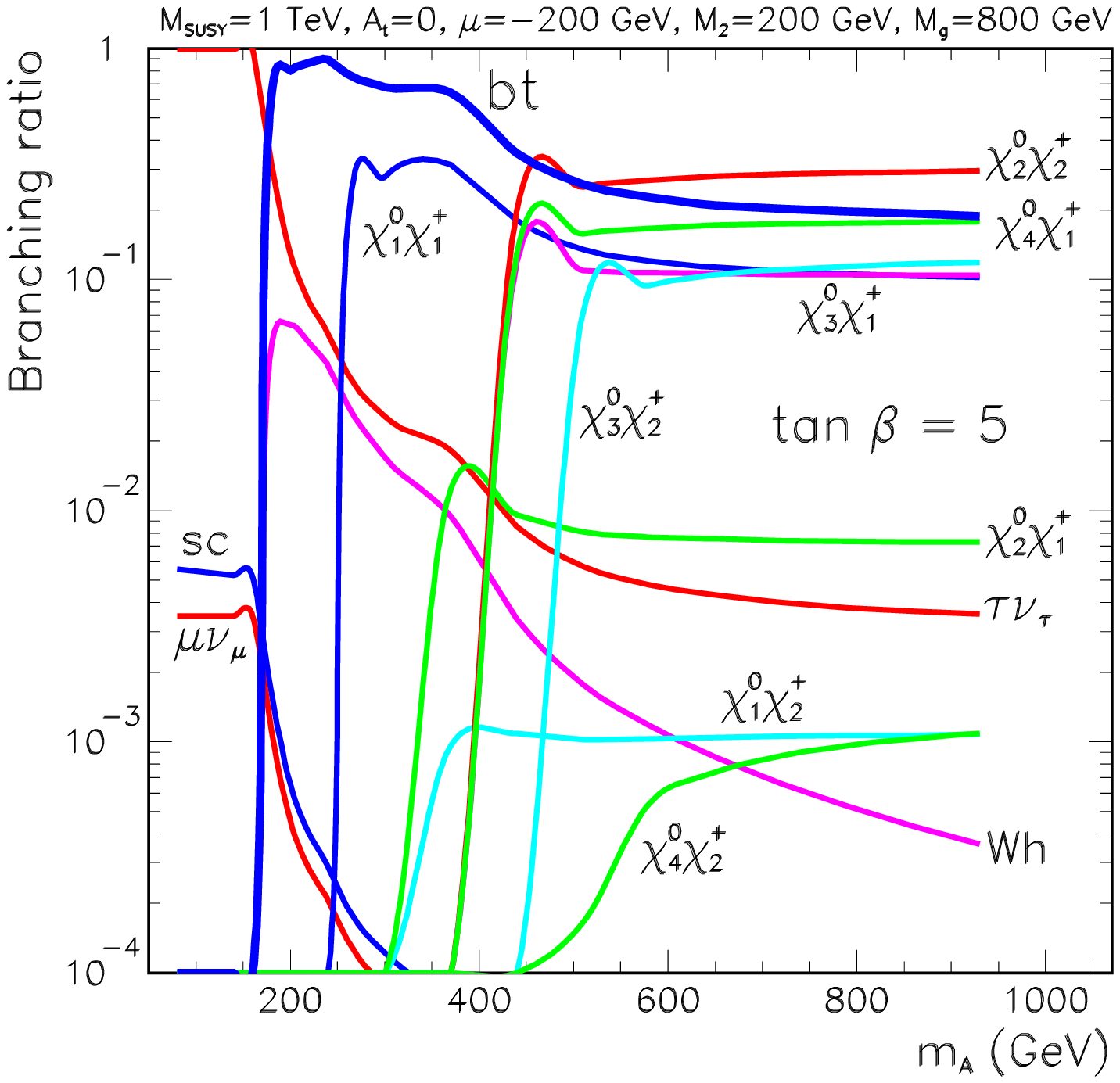,width=.65\linewidth}
\caption{The branching ratios of $H^{\pm}$ decay for $\tan \beta = 5$.}
\vspace{9mm}
\label{fig:beta5br}  
\end{center}
\end{figure}

\begin{figure}[hbtp]
\begin{center}
\epsfig{file=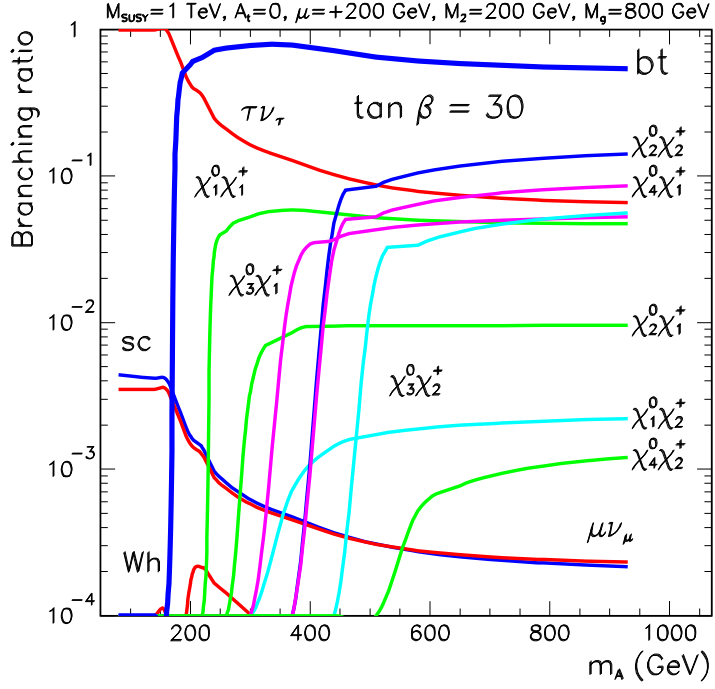,width=.65\linewidth}
\caption{The branching ratios of $H^{\pm}$ decay for
positive higgsino mass parameter $\mu$.}
\label{fig:beta30mubr}
\end{center}
\end{figure}

\begin{figure}[hbtp]
\begin{center}
\epsfig{file=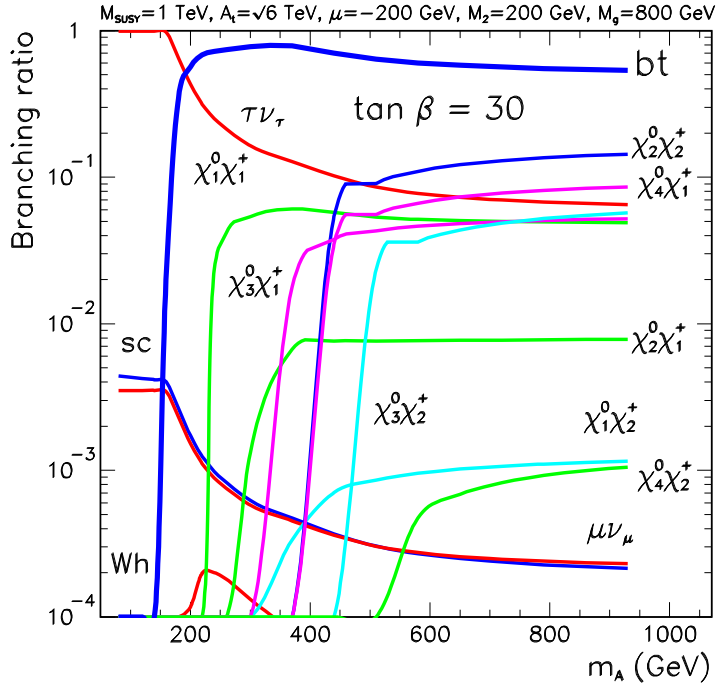,width=.65\linewidth}
\caption{The branching ratios of $H^{\pm}$ decay assuming maximal 
stop mixing, $A_{t}=\sqrt{6}$ GeV.}
\label{fig:beta30atbr}
\end{center}
\end{figure}

Only few decay channels seem to be useful for the search of the charged
Higgs boson at LHC. The most promising one is the  $H^{\pm} 
\rightarrow \tau\nu$ decay mode, in particular in the $tbH^{\pm}$ 
associated production
 with a top quark decaying to a fully hadronic final state \cite{hplus}. The $H^{\pm} \rightarrow tb$ decay channel has also been proposed as a possible discovery channel
for charged Higgs at LHC in phenomenological studies \cite{moretti}.
 A further motivation for using   
the channel $H^{\pm} \rightarrow tb$ in the search of the charged Higgs boson is shown 
in Figures~\ref{fig:beta30br},\ref{fig:beta5br},\ref{fig:beta30mubr}
and~\ref{fig:beta30atbr}. The $tb$ decay becomes abruptly the dominant mode
when the mass of the $H^{\pm}$ exceeds the sum of masses of top and bottom
quarks, and is still not so heavy that the supersymmetric decay channels 
become significant. It represents more than $60\%$ to the branching 
ratio for $200$ GeV $\lesssim m_{A} \lesssim 400$ GeV, independently 
of the value of $\tan \beta$ as shown in Figures~\ref{fig:beta30br} 
and~\ref{fig:beta5br}. In the present study the 
SUSY parameters are taken to be $A_t$ = 0 (no-mixing), $M_2$ = 200 GeV, $\mu$ = -200 GeV and $M_{\tilde{g}}$ = 800 GeV, $M_{\tilde{q},\tilde{\ell}}$ = 1 TeV~\cite{LEP}. This $H^{\pm} \rightarrow tb$ decay mode is not sensitive 
to the 
soft supersymmetry breaking parameters, at least to the sign of the higgsino mass 
parameter $\mu$ neither to stop mixing seem as 
is visible from Figures~\ref{fig:beta30mubr} and~\ref{fig:beta30atbr}. 
  
\vspace{ 3mm}
\section{Signal and Background Processes}

The charged Higgs boson in association of a top quark is produced 
through the reactions
 $gb \rightarrow tH^{\pm}$ and $gg \rightarrow tbH^{\pm}$. The cross 
section for the production with 3 tagged b-jets is the sum of the 
cross sections for these two processes subsracting the common terms 
to avoid double counting \cite{moretti}.
 Here the $H^{\pm}$ is required to decay to top and bottom quarks. 
The top quarks decay further to a bottom quark and a $W$ boson.
In order to trigger the event one of $W$'s has to decay to a lepton
and its antineutrino whilst the other one is assumed to decay hadronically
 to light quarks. 

The main backgrounds are from $t \bar{t} +$ jets with two real b-jets and a
mistagged non-b-jet, and  the irreducible background due to $t \bar{t} b \bar{b}$. Potentially, there are also backgrounds
from other processes, 
 but these are much smaller. For example the process $gq \rightarrow qW$ may cause
 some background events, but is unlikely to generate 
 five jets which would fulfil the selection criteria. 
 These other backgrounds are assumed to be negligible and are not 
included in this study.

We assume an integrated luminosity of 30 fb$^{-1}$ and 
superimpose a pile-up of 3 minimum bias events on the 
physics events. For the signal this leads to the production 
(including branching ratios, without any selection cuts) of
 approximately 43~000 events 
 for $m_{A}=250$ GeV and $\tan\beta=30$, 
but only slightly over 11~000 events for $m_{A}=400$ GeV and $\tan\beta=30$. 
The expected number of background events with this luminosity 
is about~$7\times10^{6}$.

\vspace{ 3mm}
\section{Event Reconstruction}

The event generation is carried out with sPythia~\cite{Sjostrand,Mrenna} 
for both signal events using the process 
 $gb \rightarrow tH^{\pm}$ (Pythia subprocess 161) 
and the $t \bar{t}$ background (processes 81 and 82) where additional jets 
come from parton showering. Detector performance is simulated using 
CMSJET ~\cite{cmsjet} with parametrized track reconstruction performance 
based on GEANT simulations~\cite{fatsim} for b-tagging.

As a first selection criterion for an event, we require an isolated lepton with 
transverse momentum $p_{t} > 15$ GeV and at least five jets with 
$E_{t} > 20$ GeV and $ \mid \eta \mid < 2.4$ . 
Furthermore three of those jets have to be identified as b-jets 
(b-tagging is discussed in more detail in the following). 
If these conditions are not fulfilled, the event is rejected. 

\begin{figure}[hbtp]
\begin{center}
\epsfig{file=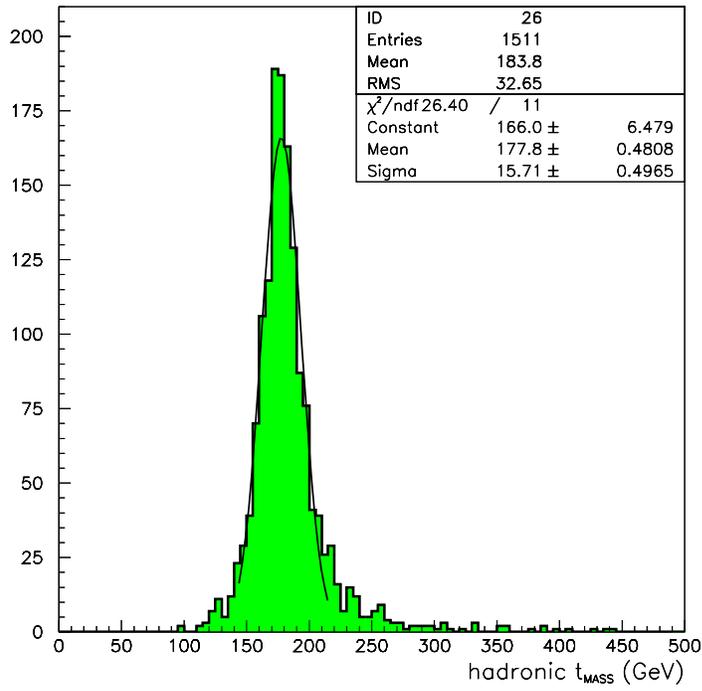,width=0.65\linewidth}
\caption{The invariant mass distribution of the hadronically decayed top quark.}
\label{fig:amtop_rec}
\end{center}
\end{figure}

\begin{figure}[hbtp]
\begin{center}
\epsfig{file=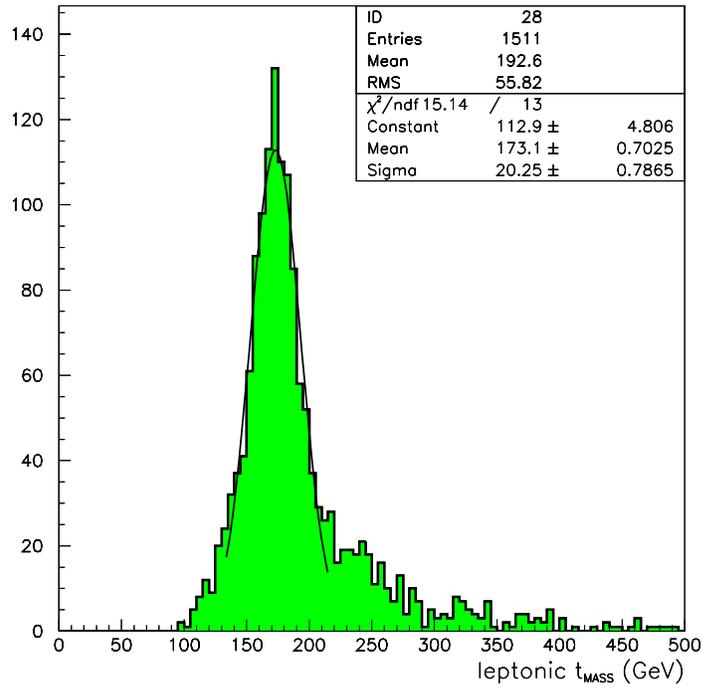,width=0.65\linewidth}
\caption{The invariant mass distribution of the leptonically decayed top quark.}
\label{fig:am_top2}
\end{center}
\end{figure}

\begin{figure}[hbtp]
\begin{center}
\epsfig{file=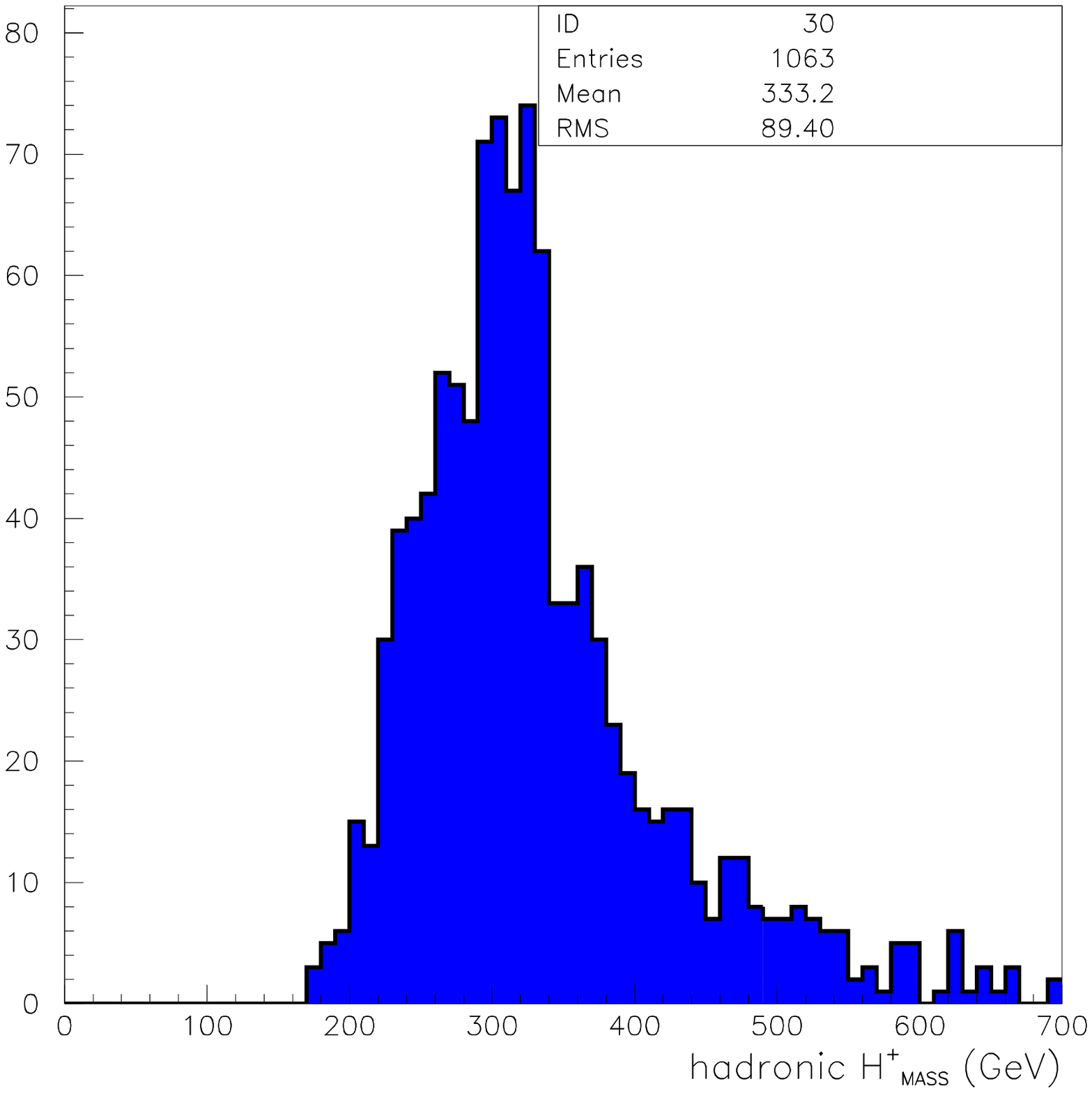,width=0.65\linewidth}
\caption{The invariant mass distribution for $H^{\pm}$ 
($m_{H^{\pm}}$=300 GeV, $tan\beta$=30) reconstructed 
from the hadronically decaying top and a b-jet.}
\label{fig:hplus1}
\end{center}
\end{figure}

\begin{figure}[hptb]
\begin{center}
\epsfig{file=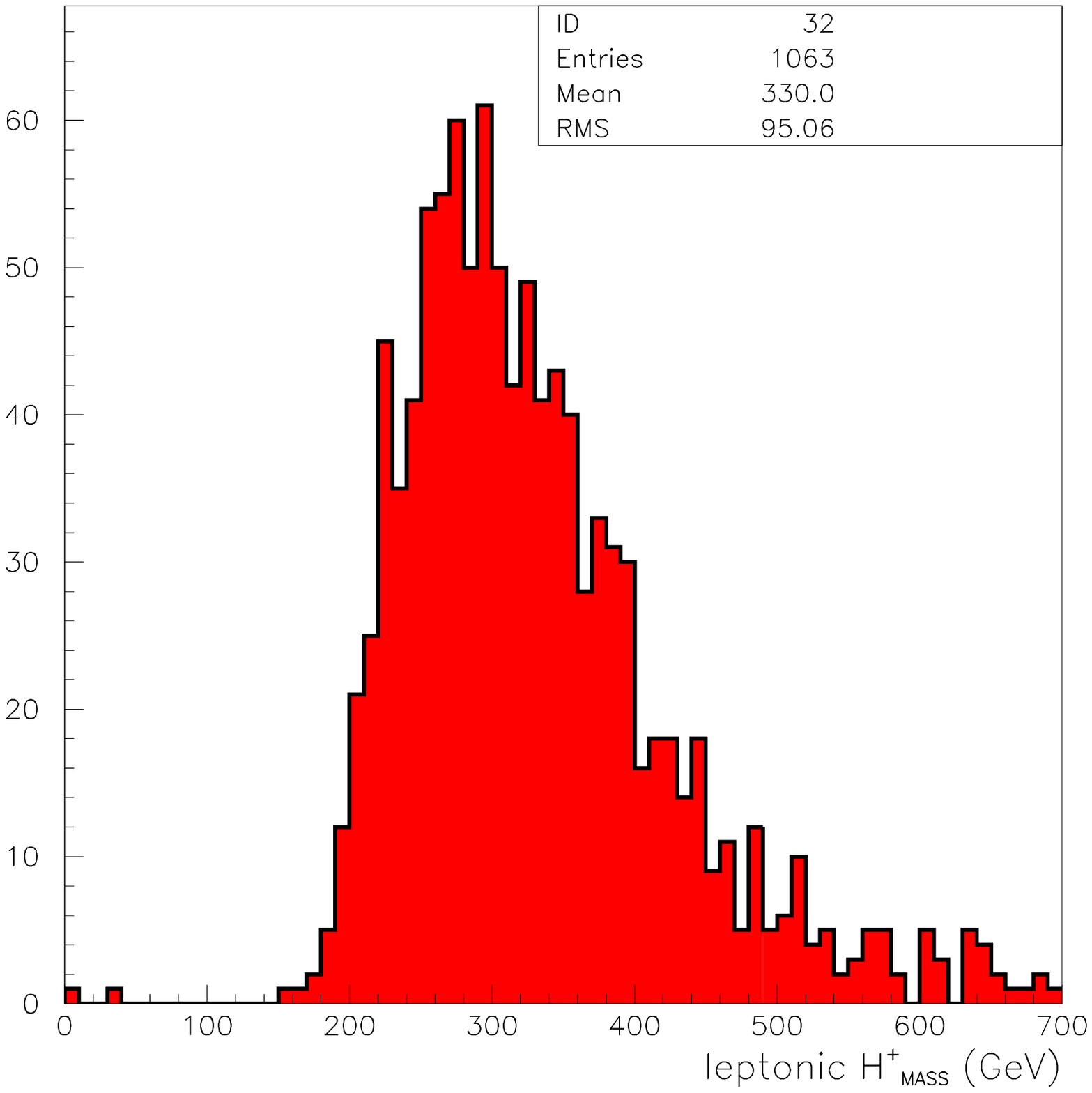,width=0.65\linewidth}
\caption{The invariant mass distribution for $H^{\pm}$
 ($m_{H^{\pm}}$=300 GeV, $tan\beta$=30) reconstructed 
from the leptonically decaying top and a b-jet.}
\label{fig:hplus2}
\end{center}
\end{figure}

Then the event reconstruction is performed as follows. 
A combination of two non-b-jets with the mass closest to the nominal
$W$ mass is chosen to be assigned to the hadronically decayed $W$ boson.
 If the reconstructed mass differs 
more than $20$ GeV form the nominal value, the event is rejected. 
Then a b-jets is chosen to reconstruct 
the mass of the top quark together with the hadronically decayed $W$,
selecting the combination with the invariant mass closest to the 
nominal top mass. 
The reconstructed hadronic top mass distribution is shown in 
Figure~\ref{fig:amtop_rec}. The other top is reconstructed from the 
leptonically decayed $W$ and a b-jet. The longitudinal component 
of the neutrino momentum is fixed using $W$ mass constraint. 
That procedure leads to two possible solutions, the smaller of which is taken. 
The missing transverse momentum is assumed to be due to the neutrino. 
The reconstructed $W$ is paired with another potential b-jet chosen to
give the mass closest to the nominal top mass to reconstruct 
the leptonic top. The mass distribution for the leptonic top is shown in Figure~\ref{fig:am_top2}.
If the mass of either of the reconstructed top quarks deviates more than 
$45$ GeV from $175$ GeV, the event is rejected. 
Finally the most probable b-jet among the remaining b-jet candidates 
is chosen to 
be paired with either the hadronic or leptonic top to reconstruct the $H^{\pm}$. 
The hadronically and leptonically reconstructed Higgs mass 
distributions are shown in Figures~\ref{fig:hplus1} and~\ref{fig:hplus2}, 
respectively. As it is not possible to know which of the two top quarks
originates from the Higgs decay this procedure leads to a large 
combinatorial background from signal events. 
The following results shown are for $m_{A} = 300$ GeV and $\tan \beta = 30$ with 
 b-likeness $>3$, unless otherwise stated. 

The background events are found to be kinematically 
very similar to the signal events. One might expect 
that extra jets accompanying $t \bar{t}$ would be considerably softer compared 
to signal events, and hard kinematical cuts on transverse momenta of 
jets would improve considerably the background rejection. 
We argue that this is not the case, as the events fulfilling 
the selection criteria for at least five jets  are in 
the rather energetic tail. 
They seem to be even harder on the average when compared to signal events 
in case of the relatively light Higgs ($m_{A} \lesssim 250$ GeV). 
For heavier Higgs ($m_{A} \simeq 400$ GeV) the jets from signal 
events are more energetic than those coming from the background, 
but not significantly. Since the total number of events in this case is 
rather small (the cross-section diminishes with increasing $m_{A}$),  
kinematical cuts, which inevitable reduce the signal as well, 
do not lead to any higher value of the formal significance. 
For these reasons we do not introduce any kinematical cuts in 
addition to those mentioned earlier and coming from the basic
detector performance. In conclusion, the severe complication 
in background rejection is due to the fact that the background very much mimics 
the shape of the signal. A possible improvement in the background rejection
could be expected from from the application of neural network methods, 
but this approach is not justified at this stage.

\vspace{ 3mm}
\section{Procedure for b-tagging and Results}

\begin{figure}[hptb]
\begin{center}
\epsfig{file=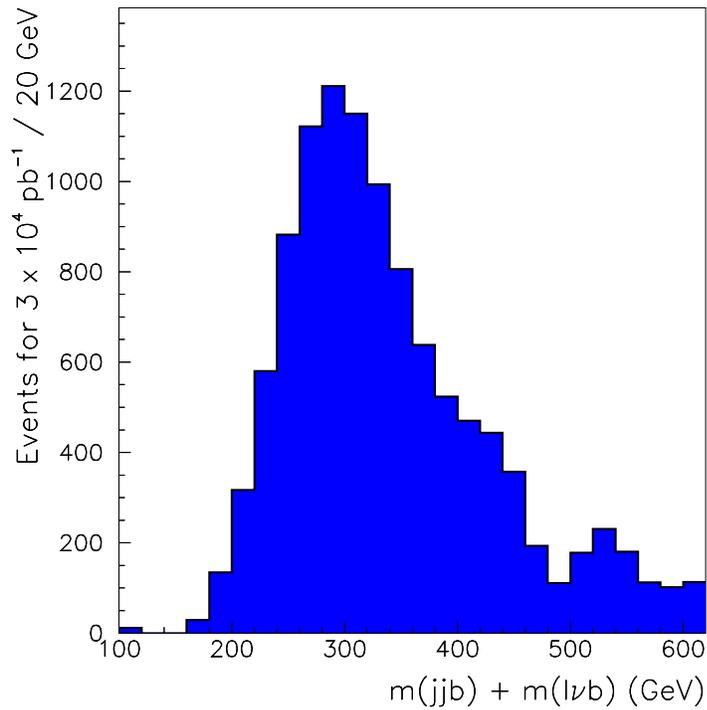,width=0.65\linewidth}
\caption{The invariant mass distribution for the background assuming
ideal b-tagging.}
\label{fig:atlas}
\end{center}
\end{figure}

The kinematical difference between the signal and the 
background being marginal, the fact that signal events contain 
one b-jet more than the background is of crucial importance. 
Actually, there is also an associated fourth b-jet in the signal event 
from the  $tbH^{\pm}$ final state and a requirement for 
four tagged b-jets would reduce the number of background events drastically. 
Unfortunately, this fourth b-jet is on the average rather soft and 
its detection is very problematic and can not therefore be used for
background reduction. On the other hand, there is an irreducible 
$t \bar{t} b \bar{b}$ background, 
so that even conceptually signal and background events may be 
indistinguishable by the number of b-jets (see also Figure~\ref{fig:atlas}). 
Thus even perfect b-tagging may be insufficient to ensure 
observation if the number of signal events is too low.

\begin{figure}[hptb]
\begin{center}
\epsfig{file=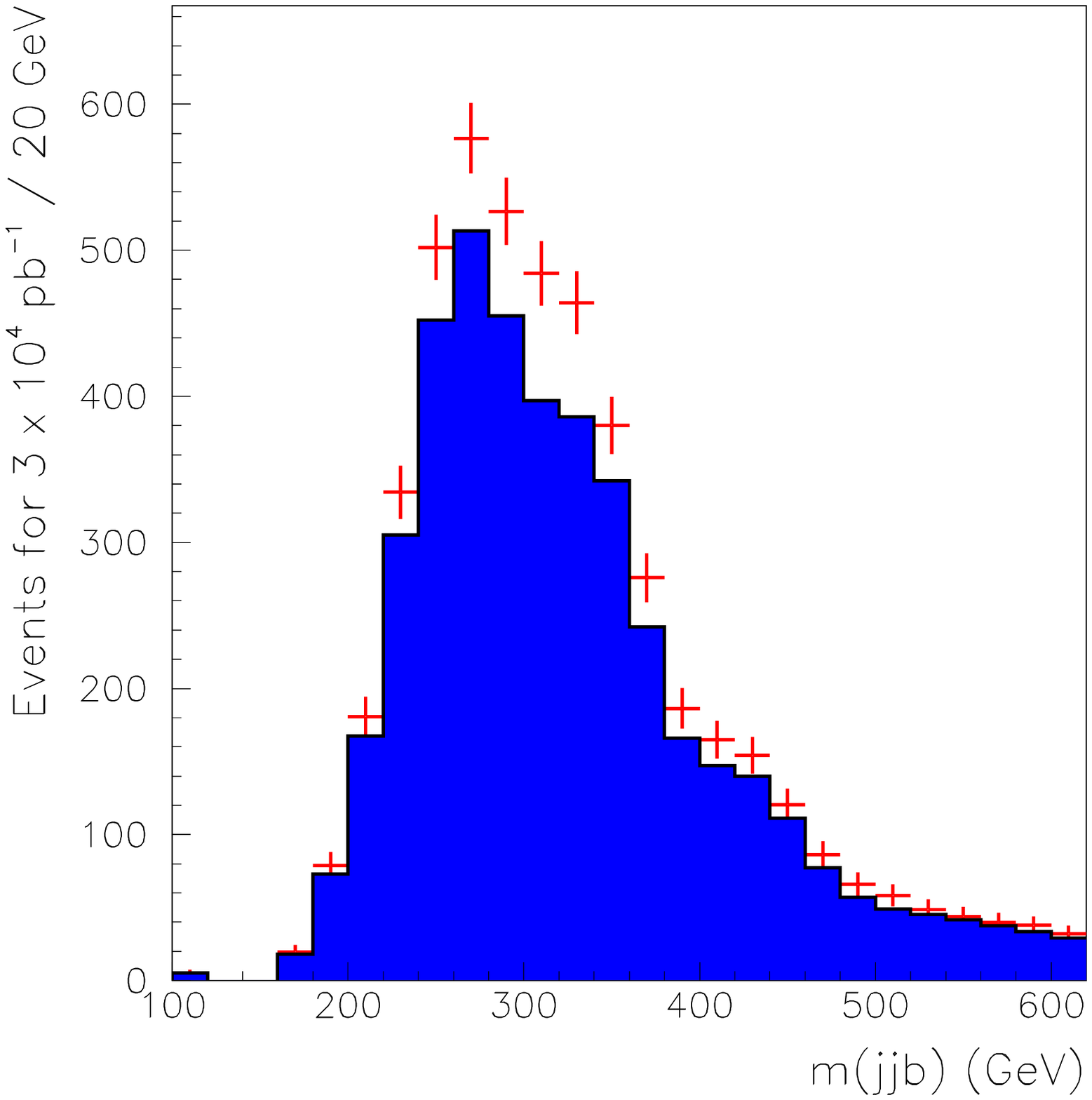,width=0.65\linewidth}
\caption{The invariant mass distribution for $H^{\pm}$ 
($m_{H^{\pm}}$=300 GeV, $tan\beta$=30) reconstructed 
from the hadronically decaying top and a b-jet superimposed on the corresponding
background for 30 fb$^{-1}$.}
\label{fig:sign_hadr}
\end{center}
\end{figure}

\begin{figure}[hptb]
\begin{center}
\epsfig{file=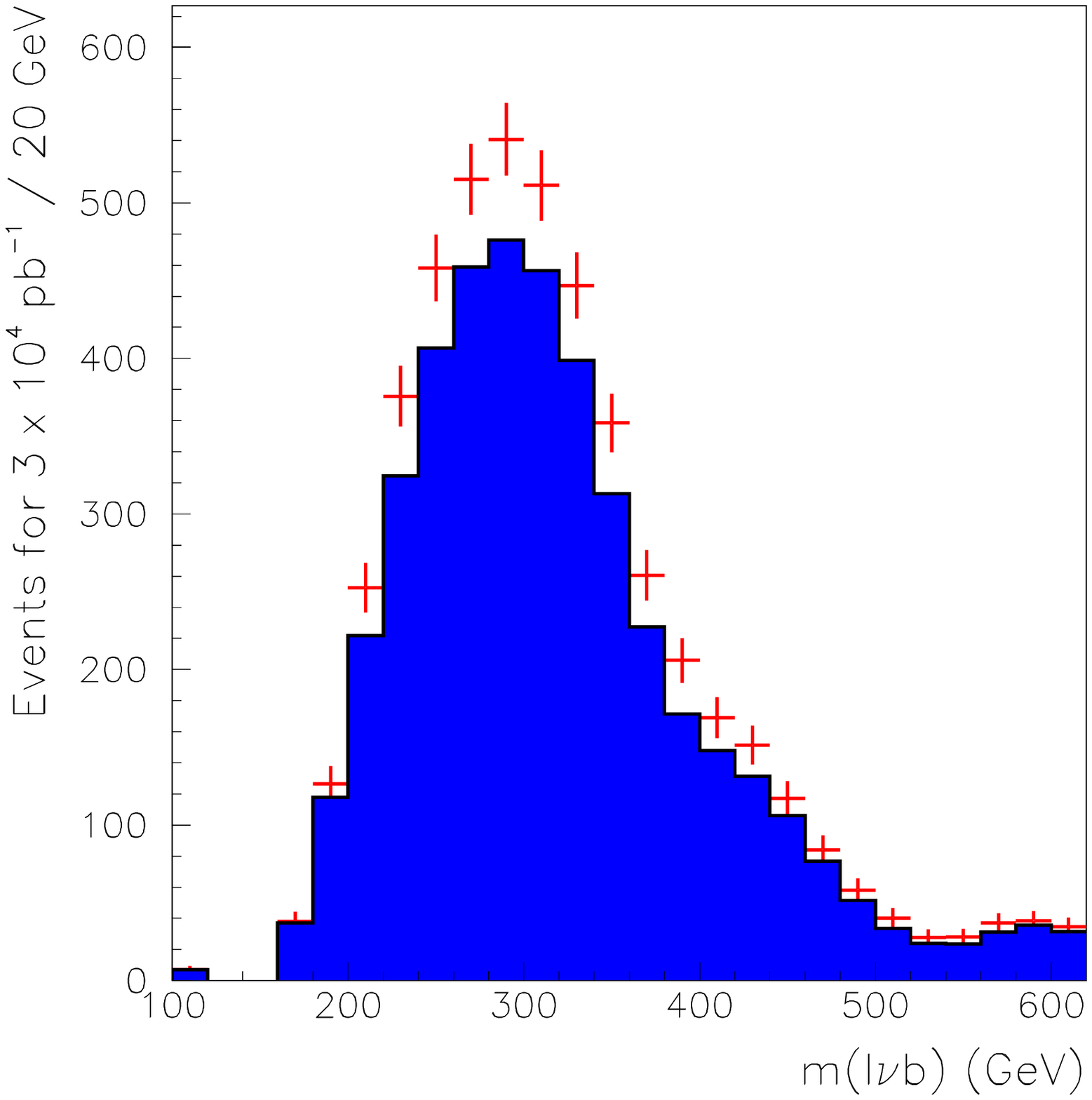,width=0.65\linewidth}
\caption{The invariant mass distribution for $H^{\pm}$
 ($m_{H^{\pm}}$=300 GeV, $tan\beta$=30) reconstructed 
from the leptonically decaying top and a b-jet superimposed on the corresponding
background for 30 fb$^{-1}$}
\label{fig:sign_lep}
\end{center}
\end{figure}

\begin{figure}[t]
\begin{center}
\epsfig{file=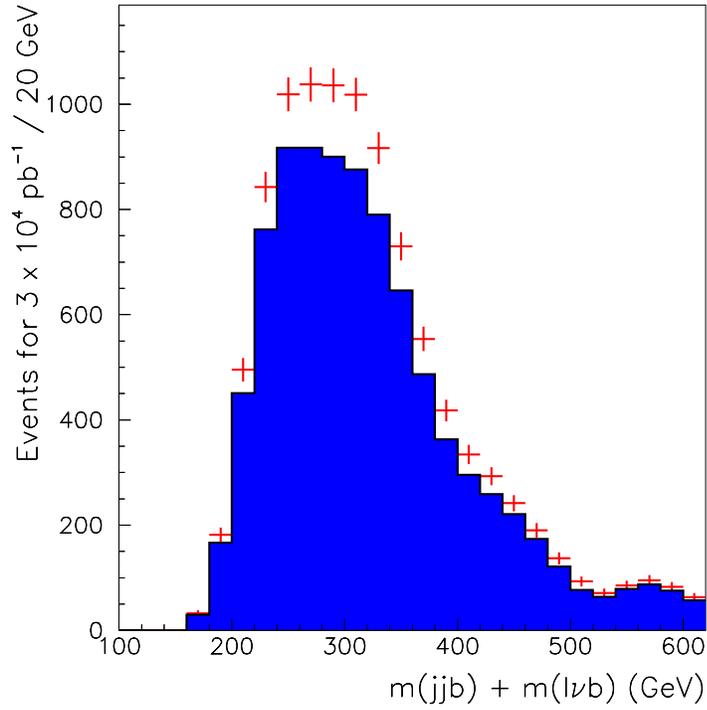,width=0.65\linewidth}
\caption{The invariant mass distribution for $H^{\pm}$ 
($m_{H^{\pm}}$=300 GeV, $tan\beta$=30) including all 
signal events (reconstructed either from hadronically or leptonically decaying top) superimposed on the total
background for 30 fb$^{-1}$}
\label{fig:sign_had_lep}
\end{center}
\end{figure}

\begin{figure}[b]
\begin{center}
\epsfig{file=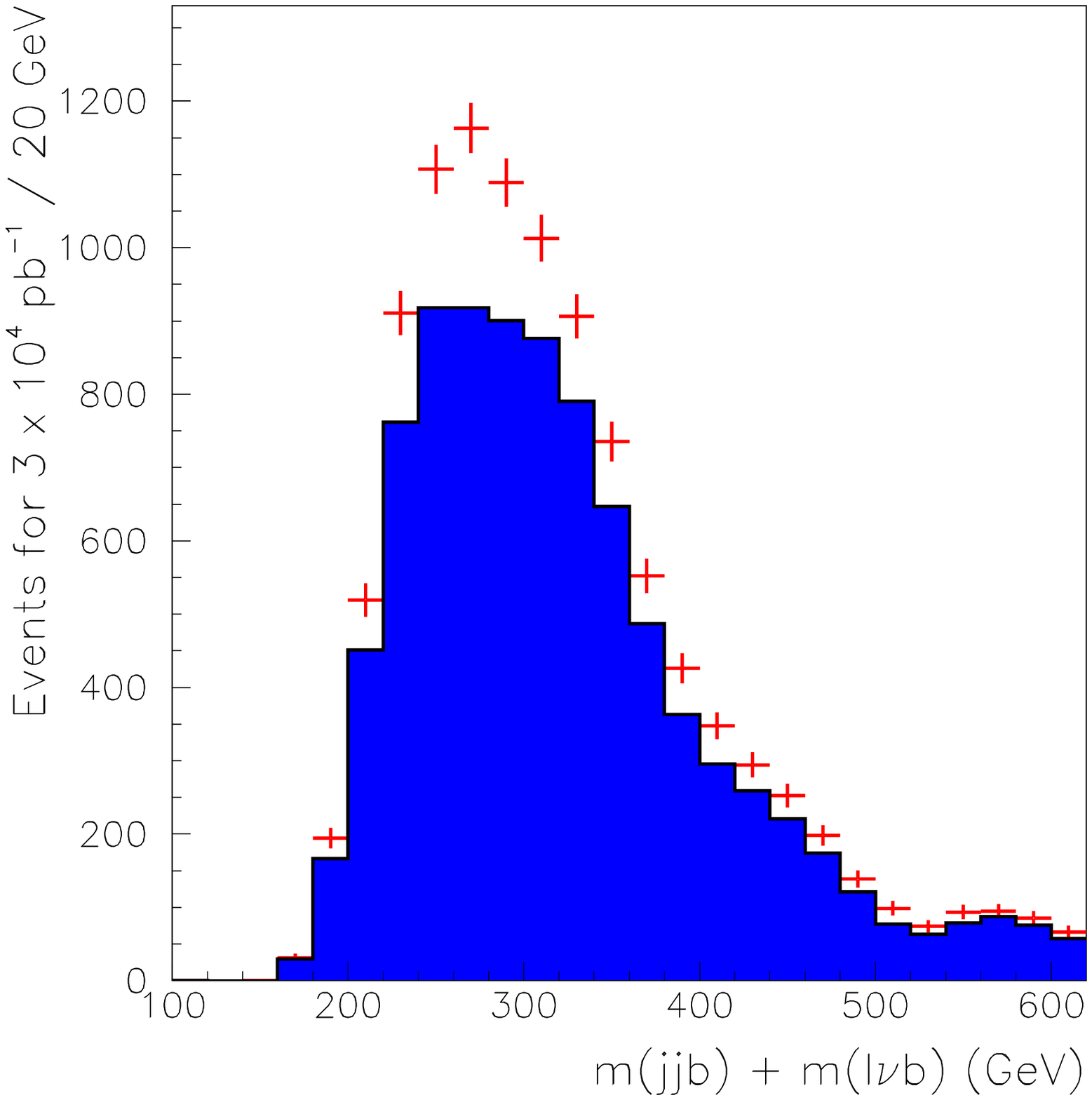,width=0.65\linewidth}
\caption{The entire signal superimposed on the total
background for $m_{A}=250$ GeV and $\tan \beta =30$.}
\label{fig:ma250}
\end{center}
\end{figure}

\begin{figure}[hptb]
\begin{center}
\epsfig{file=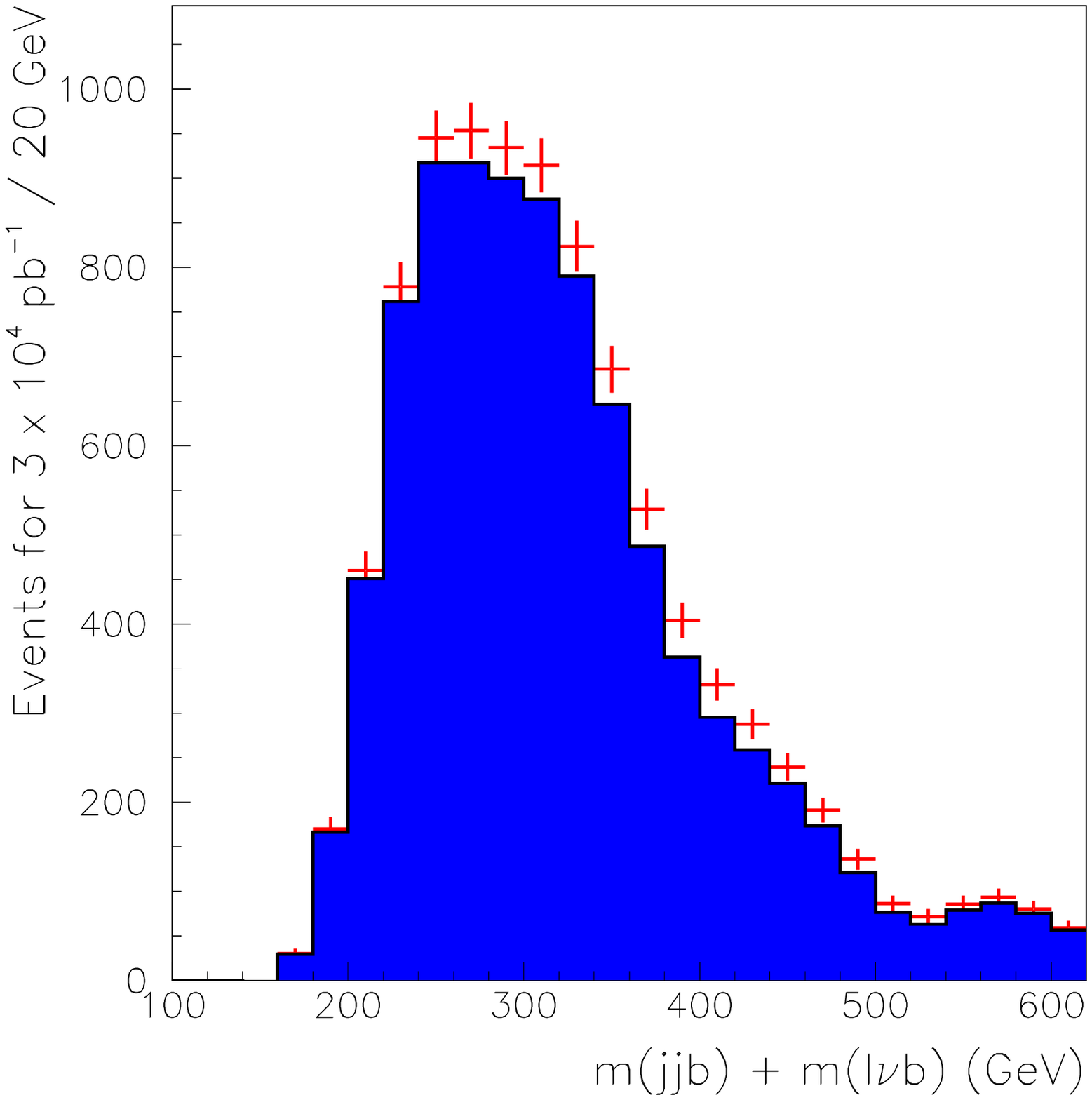,width=0.65\linewidth}
\caption{The entire signal superimposed on the total
 background for $m_{A}=400$ GeV and $\tan \beta =30$.}
\label{fig:ma400}
\end{center}
\end{figure}

The b-tagging procedure that we have used in this study has been proposed 
in~\cite{Volker}. At least two tracks in a jet cone ($\sim 0.4$) are 
required, each having transverse momentum $p_{t} > 0.9$ GeV. 
For these tracks we define the impact parameter significance $\sigma_{ip}$ 
as the ratio 
of the reconstructed transverse impact parameter to its error. 
The b-likeness of a jet is 
the significance $\sigma_{ip}$ of the second best track. 
Practically, b-jets tend to have b-likeness bigger than 2 while 
other jets seldom have b-likeness that exceeds 2. The optimal choice for a 
cut in b-likeness is found to be 3 and all the following results 
are obtained using this cut, unless otherwise stated. The distribution of the 
invariant Higgs boson mass from hadronically 
and leptonically reconstructed signal events are shown separately 
superimposed on the corresponding backgrounds in Figures~\ref{fig:sign_hadr} 
and~\ref{fig:sign_lep}. The total signal containing both hadronically 
and leptonically reconstructed top quarks together with the background 
is shown in Figure~\ref{fig:sign_had_lep}. Figures~\ref{fig:ma250} 
and~\ref{fig:ma400} show the signal superimposed on the background  
for $m_{A}=250$~GeV and $m_{A}=400$~GeV with $\tan \beta = 30$. 
The number of events decreases with increasing $m_{A}$ 
making the signal visibility difficult for $m_{A}=400$~GeV.  

\begin{figure}[hptb]
\begin{center}
\epsfig{file=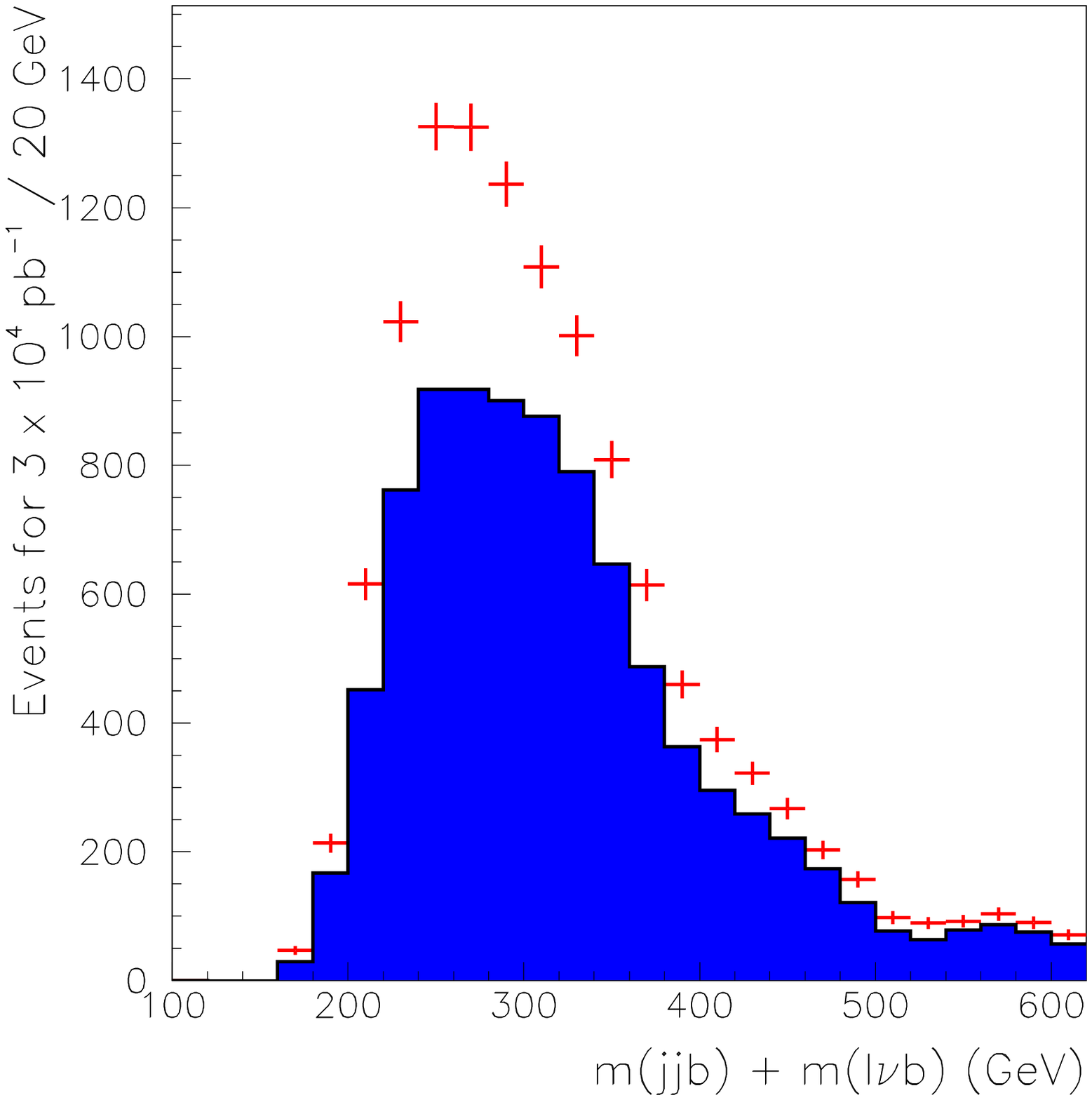,width=0.65\linewidth}
\caption{The entire signal for $m_{A}=250$ GeV 
and $\tan \beta =40$ superimposed on the total 
background with b-likeness $>3$ (significance $\sigma \simeq 21$)}
\label{fig:b3ma250tanb40}
\end{center}
\end{figure}

\begin{figure}[t]
\begin{center}
\epsfig{file=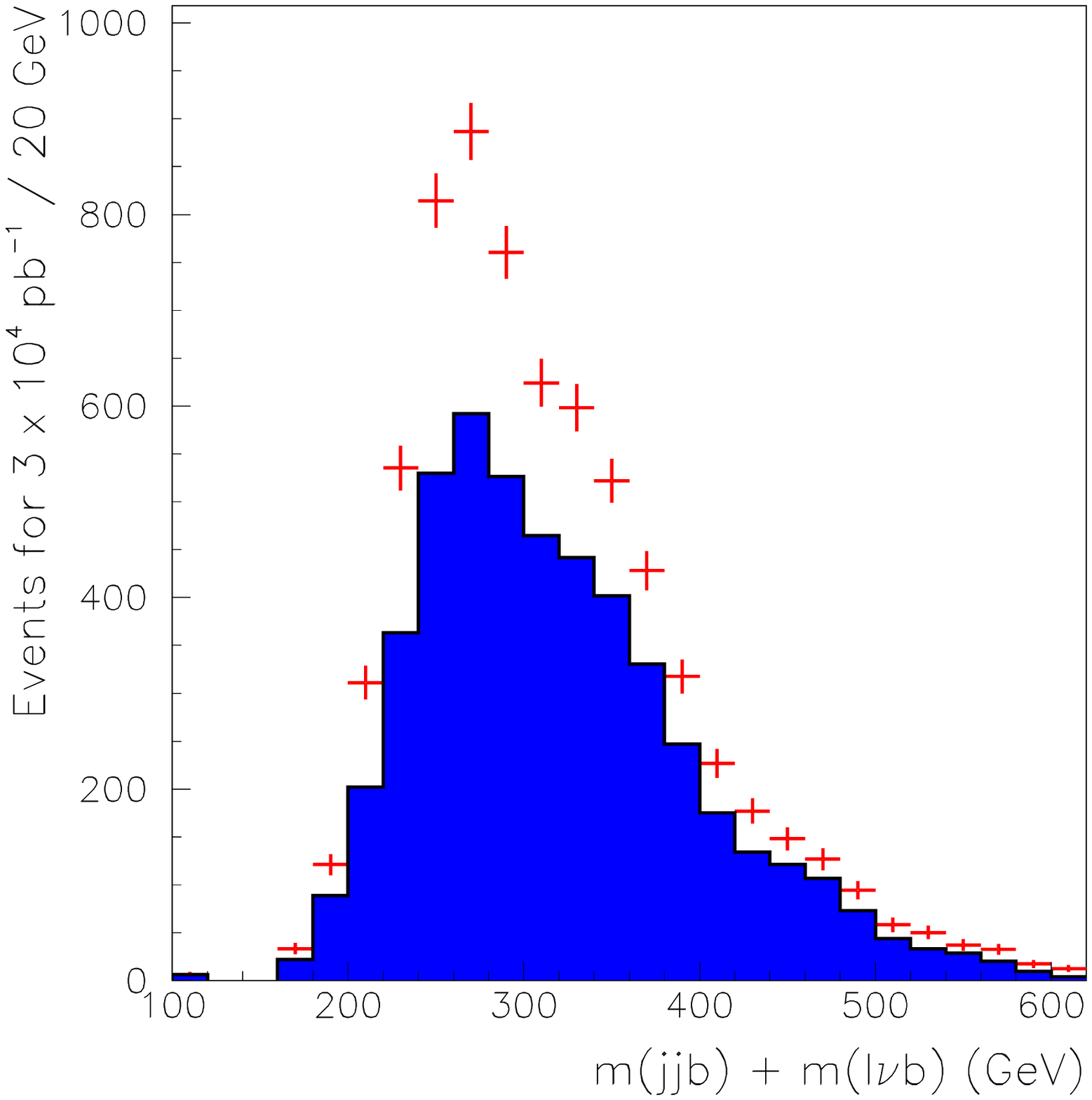,width=0.63\linewidth}
\caption{The entire signal for $m_{A}=250$ GeV 
and $\tan \beta =40$ superimposed on the total 
background with b-likeness $>4$  
(significance $\sigma \simeq 19$)}.
\label{fig:b4ma250tanb40}
\end{center}
\end{figure}

\begin{figure}[b]
\begin{center}
\epsfig{file=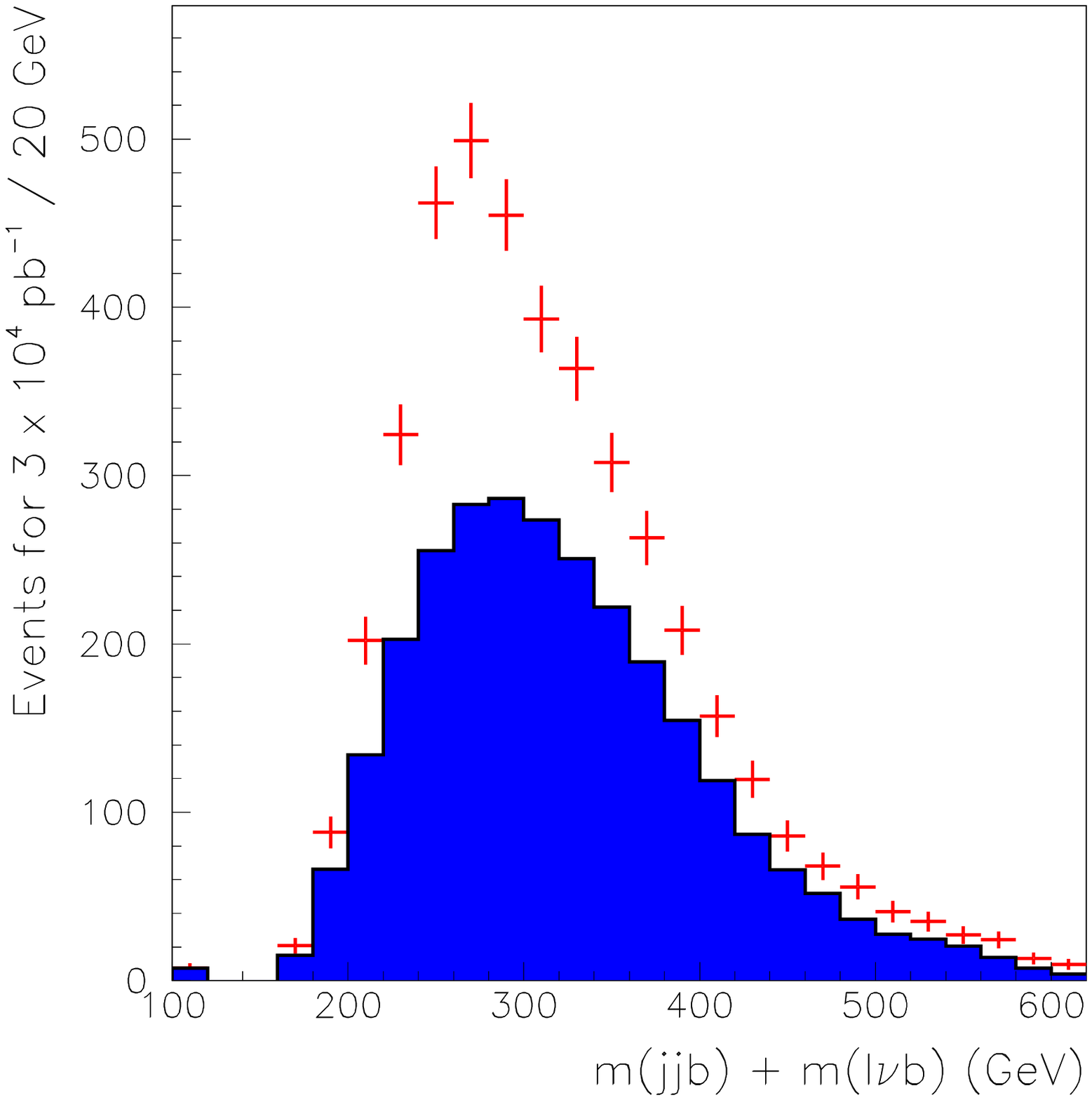,width=0.63\linewidth}
\caption{The entire signal for $m_{A}=250$ GeV 
and $\tan \beta =40$ superimposed on the total 
background with b-likeness $>5$  
(significance $\sigma \simeq 18$).}
\label{fig:b5ma250tanb40}
\end{center}
\end{figure}

\begin{figure}[hptb]
\begin{center}
\epsfig{file=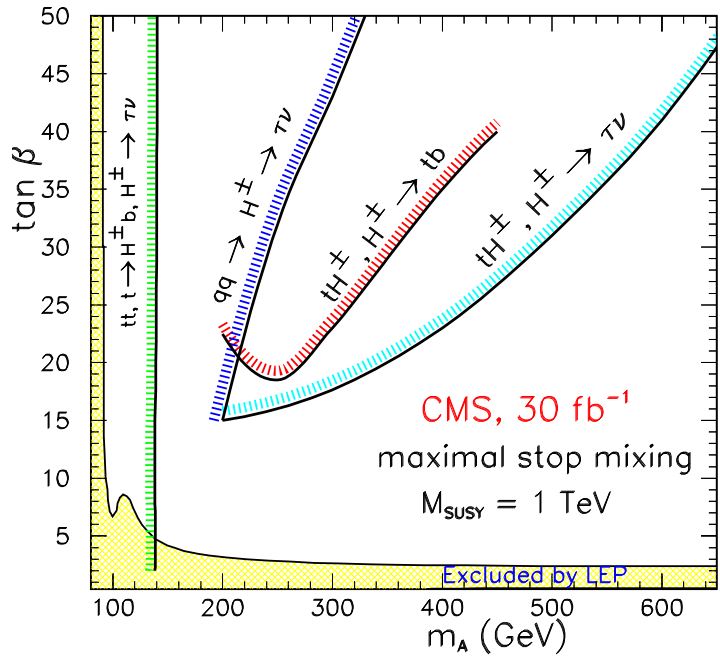,width=0.65\linewidth}
\caption{$5\sigma$-discovery contour curves for charged Higgs in CMS for 30 fb$^{-1}$.}
\label{fig:5sigma_100fb}
\end{center}
\end{figure}

The effect of the b-likeness cut is shown for $m_{A}=250$~GeV 
and $\tan \beta = 40$ in Figures~\ref{fig:b3ma250tanb40},
~\ref{fig:b4ma250tanb40} and~\ref{fig:b5ma250tanb40}. 
A harder cut reduces considerably the number of background events 
leading to a visible signal peak for b-likeness $>5$
as can be seen from Fig.~\ref{fig:b5ma250tanb40}.
The signal to background ratio is significantly improved, but 
the significance $\sigma = S/\sqrt{S+B}$ is lower due to reduced
signal statistics. The $5\sigma$-significance contour curve shown 
in Figure~\ref{fig:5sigma_100fb} is obtained using the cut b-likeness $>3$. 
The $5\sigma$-discovery regions for the channels 
$gg \rightarrow H^{\pm}tb$, $H^{\pm} \rightarrow \tau \nu$ \cite{hplus},
$q\overline{q}' \rightarrow H^{\pm} \rightarrow \tau \nu$ \cite{serguei}
and $tt$, $t \rightarrow H^{\pm}b$, $H^{\pm} \rightarrow \tau \nu$ \cite{manas} 
are also shown in the figure. 
The $gg \rightarrow H^{\pm}tb$, $H^{\pm} \rightarrow \tau \nu$ channel seems
 to be more profitable for the detection than 
the one studied in this work for any possible value of $m_{A}$ 
although the difference around $m_{A}=250$ GeV is not very large. 
The $t \rightarrow H^{\pm}b$, $H^{\pm} \rightarrow \tau \nu$ channel where the light $H^{\pm}$ is produced in top decays 
covers the low mass ($m_{A} \lsim$ 140 GeV) part of the ($m_{A},\tan \beta$) -parameter space almost independent on $tan\beta$.

We have tried to compare our results to the corresponding 
predictions of the Atlas collaboration~\cite{Assamagan}. The final results
for the discovery reach are similar. $5\sigma$-contour curves share the 
same peculiar shape, having a minimum at $m_{A} \approx 250$ GeV when 
the mass of Higgs is already significantly larger then that of 
top quark so that the jets 
coming from the decay process are energetic enough to be efficiently 
detected and 
the cross-section is still relatively large, guaranteeing plenty of events. 
However, the background invariant mass distribution in~\cite{Assamagan} is 
very different, with a monotonically decreasing behaviour, in contrast
to our result with background peaking in the signal region. The 
monotonically decreasing background shape
greatly favours the visibility of the signal peak and makes the 
determination of the background contribution under the signal peak easier. 
The kinematical cuts used in~\cite{Assamagan} are similar to ours,
thus the only possible explanation may come from the b-tagging procedure.
We tried to repeat the analysis  
using the idealised procedure for b-tagging.
Namely, we accept a b-jet with a fixed 
probability independent of its momentum and "mistag" a non-b-jet as a 
b-jet also with fixed probability. An example of the Standard Model (SM) 
background obtained this way is shown in Figure~\ref{fig:atlas} where, 
naturally after necessary kinematical cuts, all the b-jets are 
identified correctly (b-tagging efficiency = 1) and no other jets are
 mistagged as a b-jet (impurity = 0). 
Despite this, we were not able to reproduce the desirable monotonically 
decreasing background shape.
 
\vspace{ 3mm}
\section{Conclusions}

In summary, we have studied the possibility to detect a massive 
charged Higgs boson in the CMS experiment at LHC when the $H^{\pm}$ 
decays to top and bottom quarks in the process $pp \rightarrow tH^{\pm}+X$. 
This channel is found to be a potential Higgs discovery channel at LHC for
a significant part of the MSSM parameter space,
although the large background from $t\overline{t}jj$ and $t\overline{t}b\overline{b}$ events and
from signal combinatorics makes the extraction of the signal difficult.
 Because the 
decay $H^{\pm} \rightarrow tb$ does not lead to a well visible peak 
superimposed 
on the background, a precise theoretical estimate of the number of 
the SM background events is needed for the observation. 
Extraction of Higgs signal in this channel requires excellent b-tagging; 
both true b-jets have to be identified with high efficiency and 
the probability to mistag a non-b-jet must be very low; the 
$E_t$ threshold for b-jets must be low ($\sim$ 20 GeV)  
to be able to improve the separation of the signal from the expected 
number of events. A significant 
improvement can be expected at high Higgs masses ($m_{H^{\pm}}\gsim$ 400 GeV)
from the optimization of the kinematical cuts to take into account 
the much harder spectrum of b-jets from $H^{\pm}$ decay in this domain. 
A further improvement could be expected eventually with the implementation
of neural network techniques.

\vspace{12cm}

\end{document}